\documentclass[runningheads]{llncs}

\usepackage{graphicx}
\usepackage{amsmath}
\usepackage{amssymb}
\usepackage{mathtools}
\usepackage{arydshln}
\usepackage{booktabs}
\usepackage{mdframed}
\usepackage{adjustbox}
\usepackage{multirow}
\usepackage[caption=false]{subfig}
\usepackage[iso]{datetime}

\usepackage{tikz}
\usetikzlibrary{through}

\usetikzlibrary{positioning}
\usepackage{pgfplots}

\pgfplotsset{compat=1.18}

\usepackage{listings}
\usepackage{xcolor}

\spnewtheorem{construction}{Construction}{\bfseries}{\normalfont}

\newcommand*{\samp}{\xleftarrow{\text{\normalfont\tiny\$}}}

\begin{document}

\title{DID:RING: Ring Signatures using Decentralised Identifiers For Privacy-Aware Identity Proof\thanks{%
        This work is the result of research co-funded by Condatis Group Limited and The Data Lab. Permanent ID of this document: \texttt{6c52ab3735631b399f6ea954d3c91231}. Date: \today.
    }
}

\author{
    Dimitrios Kasimatis\inst{1}\orcidID{0009-0009-2036-426X} \and 
    Sam Grierson\inst{1}\orcidID{0000-0002-3625-6337}  \and
    William J. Buchanan\inst{1}\orcidID{0000-0003-0809-3523} \and 
    Chris Eckl\inst{2} \and
    Pavlos Papadopoulos\inst{1}\orcidID{0000-0001-5927-6026} \and
    Nikolaos Pitropakis\inst{1}\orcidID{0000-0002-3392-9970} \and
    Craig Thomson\inst{1}\orcidID{0000-0001-7320-7262} \and
    Baraq Ghaleb\inst{1}\orcidID{0000-0002-8361-0634}
}

\authorrunning{Dimitrios Kasimatis, Sam Grierson, William J. Buchanan, \emph{et al.}}

\institute{%
    Blockpass ID Lab, Edinburgh Napier University, Edinburgh, UK\\
    \email{\{d.kasimatis,s.grierson2,b.buchanan\}@napier.ac.uk} \and
    Condatis Group Limited, Edinburgh, UK\\
    \email{chris.eckl@condatis.com}
}
\maketitle
\begin{abstract}
    Decentralised identifiers have become a standardised element of digital identity architecture, with supra-national organisations such as the European Union adopting them as a key component for a unified European digital identity ledger. This paper delves into enhancing security and privacy features within decentralised identifiers by integrating ring signatures as an alternative verification method. This allows users to identify themselves through digital signatures without revealing which public key they used. To this end, the study proposed a novel decentralised identity method showcased in a decentralised identifier-based architectural framework. Additionally, the investigation assesses the repercussions of employing this new method in the verification process, focusing specifically on privacy and security aspects. Although ring signatures are an established asset of cryptographic protocols, this paper seeks to leverage their capabilities in the evolving domain of digital identities.
\end{abstract}

\section{Introduction}

The European Union (EU) has advocated a transition in identity management among its member states, aiming to implement European-wide electronic identification (eID) and identity trust services. To that effect, the EU introduced the Electronic Identification, Authentication, and Trust Services (eIDAS) regulation authority, tasked with overseeing the emerging digital identity infrastructure for cross-border engagements \cite{sharif2022}. Although this shift towards digital identities is crucial for developing online services and facilitating transactions, it raises concerns regarding privacy and security that warrant consideration when implementing any such framework \cite{lips2020}.

Typically, the private credentials of individuals are stored in their cryptographically secure digital wallets and, from there, can be referenced through the use of a Decentralised IDentifier (DID) \cite{sedlmeir2021}. DIDs constitute a notable transition in digital identity, introducing a novel category of globally unique identifiers that facilitate a verifiable, decentralised digital identity \cite{sporny2022}. Unlike conventional identifiers, DIDs remove the need for a centralised registration authority, forming an essential element of self-sovereign identity (SSI) systems. This concept is in harmony with the principles of evolving digital identity management, which reinstates control and privacy of identity data to the individual \cite{toth2019}. A DID is designed to function across decentralised networks, including various blockchains, and is resolved into DID documents. These documents encapsulate critical information such as public keys and service endpoints, which are requisite for secure cryptographic interactions \cite{sporny2022}. These methods outline the procedures for creating, reading, modifying, and revoking DIDs, ensuring each DID's uniqueness within its method's namespace \cite{mahalle2020}.

DIDs can be verified using cryptographic proofs, one of the more common ones being digital signatures. In this work, the application of ring signature algorithms is proposed as a means to anonymise the identification of DID documents.  For example, within the EU e-ID infrastructure, the EBSI (European Blockchain Service Infrastructure) blockchain \cite{tan2023verification} is used to store the public keys of citizens and trusted entities and where the associated digital wallet will hold the private key. When a citizen or trust entity wishes to sign a message, they will use their private key to produce the signature, and then this can be proven against their public key, which is held on the EBSI ledger.

Ring signatures \cite{rivest2001} allow signers to dynamically choose a set of public keys and sign messages on behalf of the set without revealing who the real signer is. Furthermore, it is impossible to check if any two signatures were issued by the same signer. The core property of the is the interest of this work is the provision of anonymity, making them useful for privacy-preserving protocols, such as e-voting, whistleblowing, and private transactions for cryptocurrencies \cite{backes2019}.  This paper leverages the capabilities of DIDs to present a use case for implementing ring signatures, as the SSI nature of DID-related architectures can benefit from the anonymised identification features of the method.

\vspace{-15 pt}
\subsection{Contributions}

The contributions of this paper can be summarised as follows:

\begin{enumerate}
    \item \textbf{DID Method with Ring Signature Verification.} The study introduces a new DID method that outlines the specifications to create a DID document that encapsulates the DIDs of the identity documents that are part of the signature ring. The new DID method is required to leverage the proposed verification method of ring signatures to enhance privacy and security in DID-related architectures.
    \item \textbf{Ring Signature DID Architecture.} The investigation examines the implementation of the proposed DID method within a DID-oriented architectural framework, focusing on the identification procedure. It further analyses the impact of this method on privacy and security, assessing its effectiveness and prospects. Further potential integration with privacy-preserving technologies, such as Zero Knowledge Proofs (ZKPs), is also considered to achieve digital identity verification without losing functionality.
    \item \textbf{Ring Signature Implementation Results.}  The study implements into the architecture the Borromean ring signature as a considered method and analyses its computational impact, drawing conclusions based on the number of DID documents on the ring.
\end{enumerate}

\vspace{-20 pt}
\subsection{Organisation}

The rest of the paper is structured as follows. Section \ref{sec:RelatedWork} examines related works in ring signatures and decentralised identifiers. In addition, it outlines the cryptographic-related research of ring signatures.
Section \ref{sec:RingDIDs} defines the proposed DID method and architecture for anonymous identification while also discussing the privacy and security considerations of the solution. Section \ref{sec:Results} includes the implementation results of a ring signature. Finally, section \ref{sec:Conclusion} summarises the paper's insights while giving suggestions for future work.

\section{Related work}
\label{sec:RelatedWork}

This section outlines some of the related work on ring signatures and distributed identifiers.

\vspace{-15 pt}
\subsection{Distributed Identifers} 

DIDs and DID methods were proposed as a decentralised identity standard by a W3C working group \cite{sporny2022}. Since their proposal, the landscape of DID methods has seen significant advancements, with various approaches proposed to enhance security, interoperability, and usability within decentralised systems.

Park and Nam \cite{park2021new} introduced a novel method for DIDs utilising an infinite one-way hash chain to improve security and facilitate key rotation. This method addresses the significant issue of identity theft and the challenge of managing multiple identifiers for a single entity. The effectiveness of their approach was demonstrated through implementations on Hyperledger Fabric and the Contiki Cooja simulator. In a related study, Alangot \emph{et al.} \cite{alangot2022decentralized} developed DID authentication protocols incorporating features for auditability and privacy, designed to identify malicious authentication attempts to prevent the association of authentication events with individual users.

Alzahrani \cite{alzahrani2020information} investigated the application of Information-Centric Networking (ICN) for the registry of DIDs and Verifiable Credentials (VCs), capitalising on the decentralised nature and efficient lookup capabilities of ICN to reduce network overhead and improved lookup times. In a comparative study, Alizadeh \emph{et al.} \cite{alizadeh2022performance} analysed the performance of permissionless blockchain and Distributed Hash Tables (DHT)-based registries for verifiable data, concluding that DHT-based systems offer better performance in large-scale environments.

Huh et al. \cite{huh2023did} conducted an analysis of the security and privacy considerations of the W3C DID standard \cite{sporny2022} and its universal resolver component. They introduced Oblivira, a design for privacy-preserving DID resolution that ensures the resolver processes requests without accessing their content, thereby enhancing users' privacy within digital identity frameworks.

\vspace{-15 pt}

\subsection{Ring Signatures}

The first ring signature scheme was proposed by Rivest \emph{et al.} \cite{rivest2001}, who used the RSA digital signature algorithm, along with hash and one-way functions, to instantiate a practical ring signature scheme. Abe \emph{et al.} \cite{abe2002} generalised the classical ring signature, resulting in a more standard approach to constructing ring signature schemes. Their paper was also the first to propose using a discrete logarithm-based signature scheme to construct ring signatures. Liu \emph{et al.} \cite{liu2004} further advanced ring signature schemes by enabling signatures to be generated associated with spontaneous groups of public keys. Dodi \emph{et al.} \cite{dodis2004} also works on spontaneous group signatures; however, their approach involves using cryptographic accumulators.

A major advancement in the practicality of ring signature schemes was the Borromean signature scheme of Maxwell and Poelstra \cite{maxwell2015}, enabling multiple signatures to be validated far more efficiently. The efficiency of the Borromean signature scheme was used by the Monero cryptocurrency to enable efficient validation of anonymous transactions \cite{noether2016}. Other approaches to efficient ring signatures were proposed more recently, using non-interactive zero-knowledge proofs to achieve $O(\log n)$-sized signatures such as in Bootle \emph{et al.} \cite{bootle2015} or Groth's \cite{groth2015} discrete logarithm based constructions. A general approach to non-interactive zero knowledge proof-based ring signatures was given by Backes \emph{et al.} \cite{backes2019} with efficient lattice-based constructions from Esgin \emph{et al.} \cite{esgin2019a,esgin2019b} and Yuen \emph{et al.} \cite{yuen2021}.

\vspace{-15 pt}
\subsection{Ring Signature Schemes}
\label{sec:RingSigSchemes}

In this section, we describe classical ring signature constructions, a generalised definition of a ring signature, and the necessary security and anonymity properties to which a ring signature should conform.

\vspace{-10 pt}
\subsubsection*{Notation.} Let $\mathbb{R}$, $\mathbb{Z}$, and $\mathbb{N}$ denote the sets of real, integer, and natural numbers, respectively. For a $q \in \mathbb{N}$, the notation $\mathbb{Z}_q$ denotes the congruence classes modulo $q$ and $\mathbb{Z}^\ast_q$ denotes the multiplicative group of order $q$ over the integers. Let $x \samp S$ denote a sampling of an element $x$ from the set $S$ independently and uniformly at random. The notation $x \in [n]$ will be used as shorthand for $x \in \{1, \ldots, n\}$. The function $\mathsf{negl} : \mathbb{N} \to \mathbb{R}$ is a negligible function if for every polynomial $f$ there are values $n, m \in \mathbb{N}$ such that for all $n > m$ it holds that $\mathsf{negl}(n) < \frac{1}{f(n)}$.

\subsubsection{Classical Ring Signature Constructions}

Rivest \emph{et al.}'s \cite{rivest2001} classical ring signatures for a set of $n$ public keys $\{ \mathsf{pk}_i \}_{1 \leq i \leq n}$ is constructed by computing $n - 1$ ``pseudo-signatures'' sequentially in a ring-like structure and then using the signer's secret key $\mathsf{sk}$ to close the ring with a ``real'' signature. The $n$ signatures form a ring signature on behalf of the set of public keys. Abe \emph{et al.} \cite{abe2002} provided the first generic construction for signatures using either the hash-and-one-way type (Type-H) or the three-move type (Type-T) \cite{yuen2021}.

The following briefly describes a Type-T signature and how the AOS signature constructions \cite{abe2002} function when built on top of Schnorr identification \cite{schnorr1990}. Type-T signatures consist of three steps three signing with a secret key $\mathsf{sk}$ and public key $\mathsf{pk} = g^\mathsf{sk}$ for a group generator element $g$ and message $m$: the commitment step $c = g^r$, the hashing step $e = H(m, c)$, and the response step $z = r - c \cdot \mathsf{sk}$. The resulting Type-T signature for a single public key is $\sigma = (e, z)$. Verification involves reconstructing the commitment from the signature, \emph{i.e.} $c' = g^z \cdot \mathsf{pk}^e$, and then hashing to check if $H(m, c') = e$.

\begin{figure}[ht]
    \centering
    \begin{adjustbox}{width=.6\textwidth}
        \begin{tikzpicture}[every pin edge/.style={<-}]
            \node [draw,dashed,circle through=(0:2.5)] (c) {};
            \foreach \i/\j in {0.15/0.85,1.15/1.85,2.15/2.85,4.15/4.85,5.17/5.85,6.15/6.85,8.15/8.85,9.15/9.85}
                \draw [<-] (\i/10*360:2.5) arc[radius=2.5,start angle={\i/10*360},end angle={\j/10*360}] node at ({(\i+\j)/10*360/2}:2.75) {$H$};
            \foreach \i/\j in {1/i,2/i-1,5/n,6/1,9/i+1}
                \node [draw,circle,fill=white,pin={{\i/10*360}:$\mathsf{pk}_{\j}$,$r_{\j}$}] at (c.{\i/10*360}) {$v$};
            \node [minimum height=4.5mm,fill=white,pin={0:$\mathsf{pk}_s$,$\mathsf{sk}$,$r_s$}] at (c.{10/10*360}) {};
        \end{tikzpicture}
    \end{adjustbox}
    \caption{The Type-T structure of a ring signature as defined by the generic AOS ring signatures schemes \cite{abe2002}. In the figure, $H$ corresponds to a collision-resistant hash function, $v$ is a cryptographic commitment function and the $r_i$s and $\mathsf{pk}_i$s for $1 \leq i \leq n$ are unique randomness inputs and public keys respectively.}
    \label{fig:ring-sig}
\end{figure}
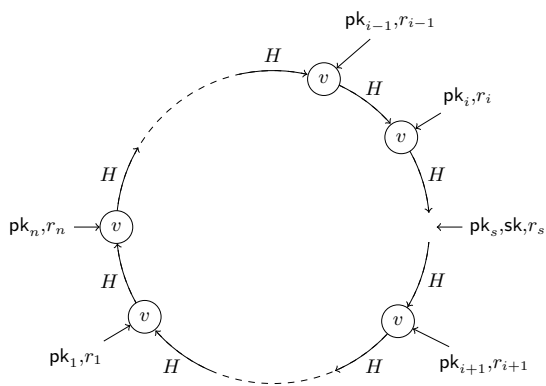
\vspace{-15 pt}

In the Type-T AOS ring signature for a ring of public keys $\mathsf{R} = \{\mathsf{pk}_i\}_{1 \leq i \leq n}$ the signer with index $s$ follows the structure in Fig. \ref{fig:ring-sig} in which the signer is assumed to have the $\mathsf{sk}$ corresponding to $\mathsf{pk}_s$. More specifically, a signer would choose randomness $r_s$ to generate their commitment and use this to compute the $i+1$\textsuperscript{th} challenge with the hash function $H$. By picking a random $i + 1$\textsuperscript{th} response $r_i$ and public key $\mathsf{pk}_i$ for $s + 1 \leq i \leq n$ then $1 \leq i \leq s - 1$ the signer can reconstruct the $i$\textsuperscript{th} commitment $c_i$ and generate the $i + 1$\textsuperscript{th} challenge $e_{i + 1}$ using the hash function $H$. This sequentially forms the ring; the final step is to then close the ring by computing $z$ from the response function. The resulting signature is $\sigma = (e_1, r_1, \ldots, r_n)$.

\subsubsection{Generic Ring Signatures}

This paper considers an adaptation of the functional definition of ring signatures given by Bender \emph{et al.} \cite{bender2006}. A ring of size $n$ is an ordered set of pubic keys $\mathsf{R} = (\mathsf{pk}_1, \ldots, \mathsf{pk}_n)$ where the notation $\mathsf{R}[i]$ denotes the $i$\textsuperscript{th} public key in the ring. Assume that the keys in the ring are unique and ordered lexicographically and that $\lvert \mathsf{R} \rvert \geq 2$, since the scheme does not intend to support standard single public key verification.

\begin{definition}[Ring Signature Scheme \cite{bender2006}]
    A ring signature scheme $\mathsf{Ring}$ is a tuple of efficient algorithms $\mathsf{gen}$, $\mathsf{sign}$, and $\mathsf{vrfy}$ defined by
    \[
        (\mathsf{pk}, \mathsf{sk}) \samp \mathsf{gen}(1^\lambda) \quad \sigma \samp \mathsf{sign}_{\mathsf{sk}, s}(m, \mathsf{R}) \quad \mathsf{vrfy}_\mathsf{R}(m, \sigma) \in \{0, 1\}
    \]
    for a security parameter $\lambda \in \mathbb{N}$ such that $\mathsf{vrfy}_\mathsf{R}(m, \mathsf{sign}_{\mathsf{sk}, s}(m, \mathsf{R}))= 1$ for any $(\mathsf{pk}_i, \mathsf{sk}_i)$ for $i \in [n]$ output by $\mathsf{gen}(1^\lambda)$, any $s \in [n]$, and any $m$ where $\mathsf{R} = (\mathsf{pk}_1, \ldots, \mathsf{pk}_n)$.
\end{definition}

A ring signature scheme should conform to the security definitions based on the basic property of anonymity, \emph{i.e.} an efficient adversary should be unable to determine which public key in a ring corresponds to the secret key used for signing, and strong unforgeability for a fixed-ring. There are some stronger definitions of anonymity, such as those considered in \cite{bender2006}; however, this paper considers signatures that correspond to a fixed ring. The following definition of anonymity considers a probabilistic security experiment in which an efficient adversary guesses a public key associated with a signature. Some previous works considered a variant of security in which the adversary is given a uniformly chosen public key in the ring $\mathsf{R}$ and asked to guess the signer with probability greater than $\frac{1}{\lvert \mathsf{R} \rvert} + \mathsf{negl}(\lambda)$ for $\lambda \in \mathbb{N}$. The signer guessing experiment is equivalent to the bit guessing experiment given below.

\begin{definition}[Signer Anonymity]
    Given a ring signature scheme $\mathsf{Ring}$ defined by $\mathsf{gen}$, $\mathsf{sign}$, and $\mathsf{vrfy}$ and an efficient adversary $\mathcal{A}$ define the following probabilistic experiment:
    \[
        \mathsf{Adv}^\text{anon}_{\mathcal{A}, \mathsf{Ring}}(\lambda) := \Pr\left[ b = b' \middle|
        \begin{aligned}
            &\mathsf{S} := (\mathsf{pk}_i)_{i \in [\mathsf{poly}(\lambda)]}\ \text{for}\ \mathsf{pk}_i \samp \mathsf{gen}(1^\lambda)\\
            &\mathcal{A}^\mathcal{O}\ \text{chooses}\ m, i_0, i_1, \mathsf{R}\subseteq \mathsf{S}\ \text{where}\ \mathsf{pk}_{i_0}, \mathsf{pk}_{i_1} \in \mathsf{R}\\
            &b \samp \{0, 1\},\ \sigma \samp \mathsf{sign}_{\mathsf{sk}_{i_b}, i_b}(m, \mathsf{R}),\ \text{and}\ b' := \mathcal{A}^\mathcal{O}(\sigma) 
        \end{aligned}
         \right]
    \]
    where $\mathcal{O}$ is a signing oracle that returns $\sigma_i \samp \mathsf{sign}_{\mathsf{sk}, s_i}(m_i, \mathsf{R}_i)$ for $i \in [q]$ where $q \in \mathbb{N}$ is the number of queries $\mathcal{A}$ can make to $\mathcal{O}$, $\mathsf{R}_i \subseteq \mathsf{S}$ and $\mathsf{pk}_s \in \mathsf{R}_i$. A ring signature scheme provides signer anonymity if $\mathsf{Adv}^\text{anon}_{\mathcal{A}, \mathsf{Ring}} \leq \frac{1}{2} + \mathsf{negl}(\lambda)$ holds for $\lambda \in \mathbb{N}$.
\end{definition}

Much like for anonymity, unforgeability has some additional stronger definitions such as those considered in \cite{bender2006}. In this paper, the ring signature schemes considered are fixed-ring schemes; therefore, the definition of security given is Strong UnForgeability against Fixed-Ring Attacks (SUF-FRA). The SUF-FRA definition given below should be familiar as if follows closely the principles of the SUF-CMA definition of security for standard digital signature schemes. The only significant difference between the two definitions of strong unforgeability is the use of a ring of public keys rather than a single uniformly generated public key.

\begin{definition}[Strong Unforgeability Against Fixed-Ring Attacks]
     Given a ring signature scheme $\mathsf{Ring}$ defined by $\mathsf{gen}$, $\mathsf{sign}$, and $\mathsf{vrfy}$ and an efficient adversary $\mathcal{A}$ define the following probabilistic experiment:
    \[
        \mathsf{Adv}^\text{suf-fra}_{\mathcal{A}, \mathsf{Ring}}(\lambda) := \Pr\left[
        \begin{aligned}
            &\mathsf{vrfy}_\mathsf{R}(m, \sigma) = 1\\
            &(m, \sigma) \notin \{(m_i, \sigma_i)\}_{i \in [q]}
        \end{aligned}
        \middle|
        \begin{aligned}
            &\mathsf{R} := (\mathsf{pk}_i)_{i \in [\mathsf{poly}(\lambda)]}\ \text{for}\ \mathsf{pk}_i \samp \mathsf{gen}(1^\lambda)\\
            &(m, \sigma) := \mathcal{A}^\mathcal{O}(\mathsf{R}) 
        \end{aligned}
         \right]
    \]
    where $\mathcal{O}$ is a signing oracle that returns $\sigma_i \samp \mathsf{sign}_{\mathsf{sk}, s_i}(m_i, \mathsf{R})$ for $i \in [q]$ where $q \in \mathbb{N}$ is the number of queries $\mathcal{A}$ can make to $\mathcal{O}$. A ring signature scheme provides unforgeability against fixed-ring attacks if $\mathsf{Adv}^\text{suf-fra}_{\mathcal{A}, \mathsf{Ring}} \leq \mathsf{negl}(\lambda)$ holds for $\lambda \in \mathbb{N}$.
\end{definition}

\section{Using Ring Signatures With Decentralised Identifiers}
\label{sec:RingDIDs}

In this section, we describe DIDs and DID methods and propose a DID method specification to enable a holder of a DID to anonymously identify themselves based on the properties of ring signature schemes.

\subsubsection*{DIDs and DID Methods.} A DID \cite{sporny2022} is a representative string that uniquely designates a digital subject, such as an individual, organisation, device, or document. The configuration and syntax of a DID is defined by the World Wide Web Consortium (W3C) standard, guaranteeing interoperability and uniformity across diverse platforms and systems. An example of a DID would be:
\begin{center}
    \texttt{did:example:bef4a730573ea233f02fbd58d83fc344}
\end{center}
where \texttt{did} is the URI scheme identifier, \texttt{example} is the DID method, and \texttt{bef4a730573ea233f02fbd58d83fc344} is the method-specific-identifier which is unique to the DID method namespace. A primary characteristic of DIDs is their capability to be resolved into DID Documents, which constitute digital records that delineate the DID subject and contain cryptographic elements like public keys and service endpoints to aid in secure communication and authentication of the DID owner's identity \cite{moreno2021}.

The decentralised aspect of a DID is enabled by distributed ledger technology, which is used to establish that DIDs are resistant to tampering, verifiable, and transferable across different systems and applications \cite{liu2020}. The decentralised approach to identity management strengthens security and privacy and gives individuals sovereignty over their digital identities, permitting them to disclose solely the information they opt for through the use of cryptographic methods such as zero-knowledge proofs \cite{yang2020}.

DID methods define how an implementer realises the features described by the specification of the DID, often associated with a specific verifiable data registry \cite{sporny2023}. One of the main advantages of using a DID method is that new DID methods can be defined to enable interoperability between implementations of the same DID method. DID methods require the definition of both a DID scheme or syntax and specific mechanisms for creating, resolving, updating and revoking DIDs and DID documents \cite{sporny2022}.

\subsection{The Ring DID Method Syntax}

The Ring DID method specification proposed in this paper is designed to conform to the DID specification currently published by the W3C Credentials Community Group \cite{sporny2022}. To implement a proposed ring signature, we leverage the verifiable credentials that are stored in a holder's digital wallet and identified by their own DIDs. The credentials' DID public keys are linked together using the newly introduced Ring DID method specification to form a DID that uses the ring signature for verification. 

The name string that will be used to identify the Ring DID method is \texttt{ring}. A DID that uses this method must begin with the prefix \texttt{did:ring}, and, per the DID specification, this string must be in lowercase. The content of the DID after the prefix for the Ring DID method is the base58 encoding of the method-specific identifier. The W3C DID specification uses the Augmented Backus–Naur Form (ABNF) to produce DIDs as URIs compliant with RFC 3986 \cite{bernerslee2005}. The proposed method's namespace-specific identifier follows this standard, which can be seen in Listing \ref{lst:abnf}. All Ring DIDs are encoded in base58 encoding using the Bitcoin/IPFS alphabets, resulting in the most alphas and digits to avoid readability issues. 

\begin{lstlisting}[float=ht,caption={The ABNF for the namespace specific identifier of the Ring DID method. All Ring DIDs are encoded in base58 with the Bitcoin/IPFS character set. Ring DIDs are 40-48 characters in length and are case-sensitive.},label={lst:abnf},basicstyle=\small\ttfamily,frame=single,captionpos=b]
ring-did = "did:ring:" idstring
idstring = 40*48(base58char)
base58char = "1" / "2" / "3" / "4" / "5" / "6" / "7" / "8"
    / "9" / "A" / "B" / "C" / "D" / "E" / "F" / "G" / "H"
    / "J" / "K" / "L" / "M" / "N" / "P" / "Q" / "R" / "S"
    / "T" / "U" / "V" / "W" / "X" / "Y" / "Z" / "a" / "b"
    / "c" / "d" / "e" / "f" / "g" / "h" / "i" / "j" / "k"
    / "m" / "n" / "o" / "p" / "q" / "r" / "s" / "t" / "u"
    / "v" / "w" / "x" / "y" / "z"
\end{lstlisting}
\vspace{-1cm}

\subsubsection*{Identifier Generator Procedure.} As per the W3C standard, each proposed DID method must specify a generation method for the method-specific identifier component of the DID. For the Ring DID method we propose the following: Let $\mathsf{R}$ be a ring of public keys such that $\lvert \mathsf{R} \rvert = n$ and the public key of the signer $\mathsf{pk}_s = \mathsf{R}[s]$ for the index $s \in \mathbb{N}$. Define a collision-resistant hash function $H$ as $H : \{0, 1\}^\ast \to \{0, 1\}^{512}$ and compute
\[
   r \samp \{0, 1\}^{\geq \lambda} \qquad H(\mathsf{pk}_s \mid\mid  r \mid\mid \mathtt{ring} || \mathsf{pk}_1 \mid\mid \cdots \mid\mid \mathsf{pk}_n)
\]
for a $\lambda \in \mathbb{N}$ where $\mathtt{ring}$ is the ASCII of the method name represented in bytes. The method-specific ID for the user is the first $256$ bits of the output of $H$ encoded in base58. For example, a valid \texttt{ring} DID may be:
\begin{center}
    \texttt{did:ring:BZEwrymg8P7aCwpJVGzuXHejijUBsmoCLWR4dgfNPuWd}.
\end{center}

\subsection{DID Operation Definitions}

The W3C DID core standard requires that DID methods define mechanisms to create, read, update, and delete a DID and its DID document. In the following, we outline each required mechanism for the \texttt{did:ring} method and provide an overview of the intended functionality of \texttt{did:ring}.

\begin{lstlisting}[float=ht,caption={A valid \texttt{did:ring} document for a ring of two public keys defined under the \texttt{LinkedDomain} services as \texttt{serviceEndpoint}s. },label={lst:document},basicstyle=\small\ttfamily,frame=single,captionpos=b]
{
    "@context": ["https://www.w3.org/ns/did/v1"],
    "id": "did:ring:IDENTIFIER",
    "authenticationMethod": [{
        "id": "did:ring:IDENTIFIER",
        "type": "RING_VERIFICATION_METHOD",
        "controller": "did:ring:IDENTIFIER",
        "publicKeyBase58": "RING"
    }],
    "service": [{
        "id": "did:DID_1:IDENTIFIER_1#cred-1",
        "type": "LinkedDomains",
        "serviceEndpoint": "did:DID_1:IDENTIFIER_1"
    }, {
        "id": "did:DID_2:IDENTIFIER_2#cred-2",
        "type": "LinkedDomains",
        "serviceEndpoint": "did:DID_2:IDENTIFIER_2"
    }]
}
\end{lstlisting}
\vspace{-1cm}

\subsubsection*{Create (Register) Method.} The create or register method initialises a \texttt{did:ring} DID and creates the DID document. To create a \texttt{ring:did}, the user must specify a minimum of $2$ credentials that will constitute the ring. In this paper, the DID method we propose is designed to use an arbitrary ring signature method, \emph{i.e.} AOS \cite{abe2002} or Borromean \cite{maxwell2015} signatures. As such, the credentials that constitute the ring signature need to use signing algorithms that are supported by the \texttt{ring:did} signing method.

Forming the actual ring of the ring signature is achieved by pointing to credentials \emph{via} \texttt{serviceEndpoint}s which are \texttt{LinkedDomains}. The \texttt{serviceEndpoints}s act as links to the DIDs corresponding to each verifiable credential encapsulated within the ring, thereby establishing a verifiable linkage. The \texttt{serviceEndpoint} parameter is essentially a DID that contains a verifiable signing key enabling ring signature's ability to generate a legitimate signatory output. Listing \ref{lst:document} shows an example DID document for the \texttt{did:ring} method for a size $2$ ring.

\vspace{-15 pt}
\subsubsection*{Read (Resolve) Method.} Resolution of a \texttt{did:ring} DID is slightly more involved that simple verification with a single public key digital signature scheme. The \texttt{did:ring} would be resolved by the DID resolver, which would take each \texttt{serviceEndpoint} specified by the \texttt{did:ring} DID document and find the public key associated with the credential those endpoints point to. The public keys found by the resolved from the ring that would be used in a ring signature verification specified by the \texttt{authenticationMethod}.

\vspace{-15 pt}
\subsubsection*{Update (Replace) Method.} A \texttt{did:ring} can be dynamically updated by invoking add and remove key mechanisms that essentially just add new endpoints for the resolver to locate public keys from. The \texttt{controller} attribute, while not mandatory, can play a role in defining the entities authorised to enact changes to the DID document.

\vspace{-15 pt}
\subsubsection*{Delete (Revoke) Method.} Due to the nature of a ring signature,  a \texttt{did:ring} DID cannot be deleted or revoked. However, by deleting or revoking the credentials that constitute the ring of public keys that is used in the signing process the \texttt{did:ring} DID would be unable to generate valid signatures.
\vspace{-15 pt}
\subsection{Ring DID Identification Architecture}

\begin{figure}
\centering
\includegraphics[width=1\textwidth]{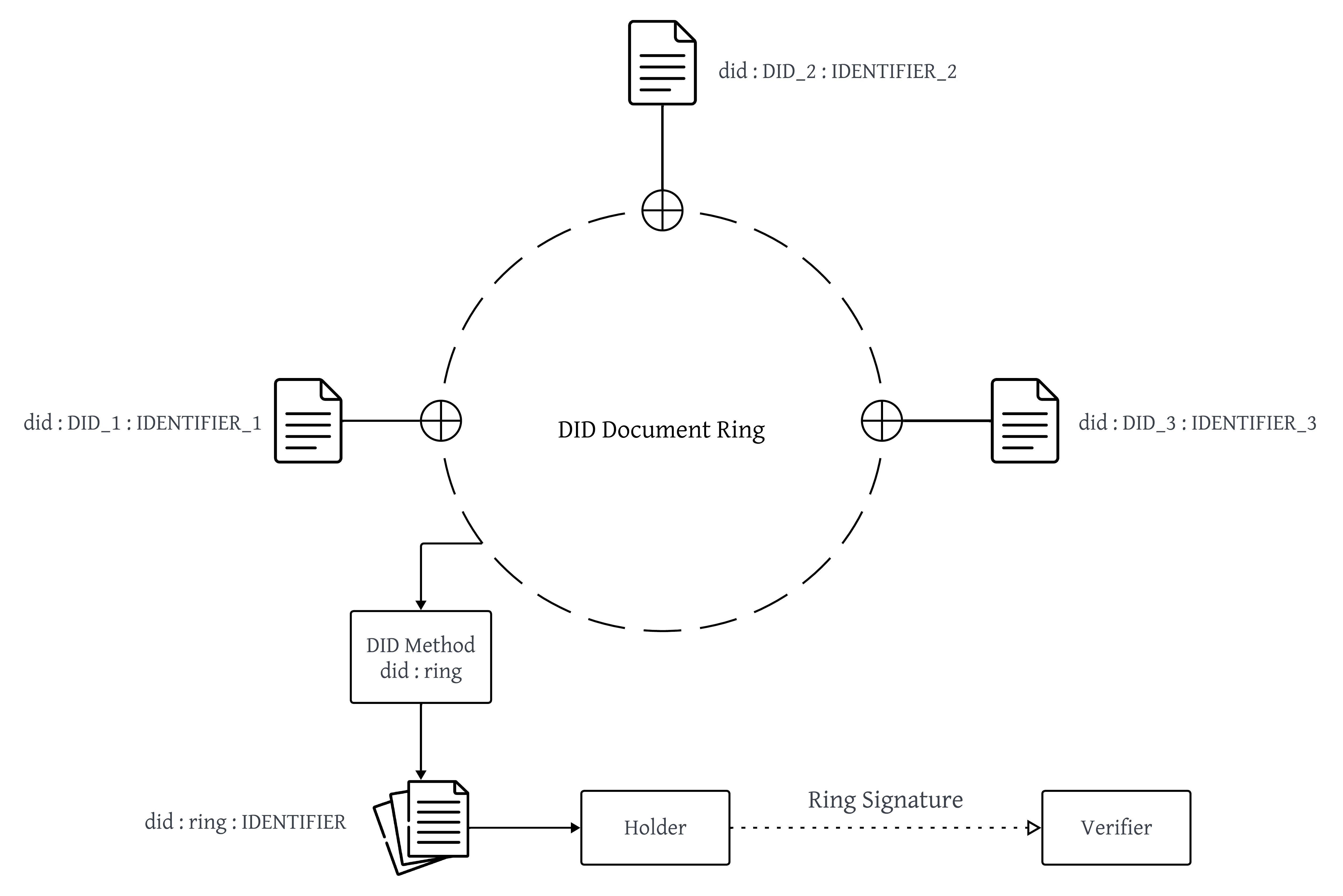}
\caption{Ring DID Identification Architecture.}
\label{fig2}
\end{figure}

The proposed DID method, referred to as \textit{did:ring}, is implemented within a DID-based framework, as depicted in Figure \ref{fig2}. The figure is a template digital identification architecture that consists of several DID documents, specifically \textit{did:DID\_1:IDENTIFIER\_1}, \textit{did:DID\_2:IDENTIFIER\_2}, and \textit{did:DID3:IDENTIFIER\_3}. These documents are pre-existing, verified credentials of the holder and constitute the elements of the \textit{ring} structure. The primary function of the \textit{did:ring} method is to aggregate these individual DIDs into a unified ring DID document, identified as \textit{did:ring:IDENTIFIER}. This document incorporates the DIDs of the verified credentials and employs ring signatures as its verification mechanism. Upon receiving a request for identification from a verifier, the holder is able to provide a signatory output from the ring DID, thereby verifying their identity without disclosing the specific DID document used in the process.

\vspace{-15 pt}

\subsection{Security and Privacy Considerations}

\subsubsection*{Security Considerations.} The implementation of the ring signature within the proposed DID method and architecture is primarily focused on reinforcing security and privacy, which are fundamental in the governance of SSI digital identities. By integrating the ring signature mechanism into the DID framework, the architecture achieves a higher level of security through the obfuscation of the DID document, which is essential for verifying the identity of the holder. According to the specifications outlined in the DID method schema, although the DIDs comprising the ring are disclosed, it remains impossible to ascertain which specific DID document is employed in the generation of the signatory output. This security through obscurity complicates potential unauthorised attempts to decipher or alter the signature, essentially making the system tamper-proof, ensuring the integrity and authenticity of the digital identities are maintained.

\subsubsection*{Privacy Considerations.} Concerning privacy, the architecture not only conceals the identity of the signing DID document but also establishes a foundational framework conducive to further advancements in privacy preservation. This is achievable through the potential integration of ZKPs, which offer a method for validating transactions or data exchanges without disclosing the underlying information. Such an enhancement would significantly augment the privacy aspect of the architecture, making it a versatile and secure foundation for managing digital identities in the realm of SSI.

\vspace{-15 pt}

\section{Implementation Results}
\label{sec:Results}

The proposed method can be used with a range of ring-based signatures. Table \ref{tab:results-1} outlines the results based on a Borromean Ring Signature method \cite{asecuritysite_40156}. This uses an AWS t3.medium instance with an Intel Xeon Platinum 8000 series processor, 3.1 GHz clock speed, and 4GB of RAM. We can see, in this evaluation, that a two DID ring can perform 5,849 ring creations, while we get 1,296 ring signing operations and 995 ring verifications. The greater the number of DIDs on the ring thus reduces the over performance of the processing.

\vspace{-15 pt}
\begin{table}[ht]
    \centering
    \caption{Borromean Ring Signature Evaluation (operations per second)}
    \label{tab:results-1}
    \begin{tabular}{llll}
        \toprule
        Ring size & Ring creation & Ring signing & Ring verification \\ 
        \midrule
        2  & 5,849 & 1,296 & 995  \\
        3  & 5,128 & 888  & 606  \\
        4  & 3,525 & 573  & 540  \\
        5  & 2,951 & 491  & 410  \\
        6  & 2,584 & 402  & 330  \\
        7  & 2,198 & 297  & 310  \\
        8  & 2,142 & 302  & 265  \\
        9  & 1,691 & 255  & 197  \\
        10 & 1,607 & 220  & 183  \\ 
        \bottomrule
    \end{tabular}
\end{table}

\section{Conclusion}
\label{sec:Conclusion}

This study addresses the integration of verification processes within digital identity frameworks utilising DIDs. It introduces a novel DID method designed to incorporate ring signatures alongside a practical case for its architectural deployment. This identification approach offers significant improvements in security and privacy by obscuring the DID document of the holder during the verification process and ensuring the integrity of the signatory output against malicious interference. The investigation proposes a distributed ledger-agnostic method, with its applicability to a functioning ledger necessitating further exploration and refinement.

Future research should focus on refining the proposed DID method. In particular, adjustments to the identifier generator procedure are recommended, such as directly incorporating the public keys from the ring DID documents rather than relying on \textit{LinkedDomains}, assuming this can be achieved without data loss. Moreover, the potential of ring signatures warrants further investigation to enhance anonymous identification capabilities and computational efficiency, particularly as the number of ring elements grows. Despite these challenges, ring signatures emerge as a promising solution for identity verification within DID-based frameworks, offering a balance between user privacy and security.

\bibliographystyle{splncs04}
\bibliography{references}

\end{document}